\documentclass[envcountsame]{llncs}

\usepackage{amsmath}
\usepackage{graphicx}
\usepackage{amssymb}
\usepackage{float}

\usepackage{tikz}

\renewenvironment{proof} 
               {\noindent \textit{\textbf{Proof.}}~ }     
               {\hfill\rule{2mm}{2mm} \vspace{\parskip} } 

\floatstyle{ruled}
\newfloat{algorithm}{htbp}{loa}
\floatname{algorithm}{Algorithm}

\begin{document}

\title{Minimum Cost Homomorphisms with Constrained Costs}
\author{Pavol Hell\inst{1}
\and
Mayssam Mohammadi Nevisi\inst{1}
}

\institute{School of Computing Science, Simon Fraser University, Burnaby, Canada,
\email{pavol,mayssamm@sfu.ca}\thanks{This work was partially supported by NSERC (Canada) and by 
ERCCZ LL 1201 Cores (Czech Republic); it was done in part while the first author was visiting the Simons 
Institute for the Theory of Computing}}

\maketitle

\begin{abstract}
The minimum cost homomorphism problem is a natural optimization
problem for homomorphisms to a fixed graph $H$. Given an input
graph $G$, with a cost associated with mapping any vertex of 
$G$ to any vertex of $H$, one seeks to minimize the sum of 
costs of the assignments over all homomorphisms of $G$ to $H$. 
The complexity of this problem is well understood, as a function 
of the target graph $H$. For bipartite graphs $H$, the problem is
polynomial time solvable if $H$ is a proper interval bigraph, and
is NP-complete otherwise. In many applications, the costs may be
assumed to be the same for all vertices of the input graph. We
study the complexity of this restricted version of the minimum
cost homomorphism problem. Of course, the polynomial cases are
still polynomial under this restriction. We expect the same will
be true for the NP-complete cases, i.e., that the complexity
classification will remain the same under the restriction. We
verify this for the class of trees. For general graphs $H$, we 
prove a partial result: the problem is polynomial if $H$ is a 
proper interval bigraph and is NP-complete when $H$ is not chordal 
bipartite.
\end{abstract}


\section{Introduction}
\label{5:sec:intro}

Suppose $G$ and $H$ are graphs (without loops or multiple edges).
A {\em homomorphism} $f : G \to H$ is a mapping $V(G) \to V(H)$
such that $f(u)f(v) \in E(H)$ whenever $uv \in E(G)$. For a fixed
graph $H$, a number of computational problems have been considered.
In the {\em homomorphism problem}, one asks whether or not an input 
graph $G$ admits a homomorphism to $H$. It is known that this problem 
is polynomial time solvable if $H$ is bipartite, and is NP-complete 
otherwise \cite{hn}. In the {\em list homomorphism problem}, the
input graph $G$ is equipped with lists (sets) $L(x) \subseteq V(H)$,
for all $x \in V(G)$, and one asks whether or not there exists a
homomorphism $f:G \to H$ with $f(x) \in L(x)$ for all $x \in V(G)$.
This problem is known to be polynomial time solvable if $H$ is an
interval bigraph, and is NP-complete otherwise \cite{fhh}. (An {\em 
interval bigraph} is a bipartite graph $H$ with parts $X$ and $Y$
such that there exist intervals $I_x, x \in X$, and $J_y, y \in Y$,
for which $xy \in E(H)$ if and only if $I_x \cap J_y \neq \emptyset$.)
In this paper we address the {\em minimum cost homomorphism problem}, 
in which the input graph is equipped with a cost function 
$c:V(G)\times V(H) \to \mathbb{N}$
and one tries to minimize the total cost $\sum_{u\in V(G)}{c(u,f(u))}$.
Minimum cost homomorphism problems were introduced in~\cite{gutin}. They 
were motivated by an application in repair and maintenance scheduling;
however, the problem arises in numerous other contexts, e.g. in the minimum 
colour sum problem and the optimum cost chromatic partition problem \cite{barnoy,supowit}.
To state it as a decision problem, the input includes an integer $k$, and
one asks whether or not there exists a homomorphism of total cost at most $k$.
This problem is known to be polynomial time solvable if $H$ is a proper interval
bigraph, and is NP-complete otherwise \cite{min-cost}. (An interval bigraph is 
a {\em proper interval bigraph} if the above two families of intervals 
$I_x, x \in X$, and $J_y, y \in Y$ can be chosen to be inclusion-free, i.e., 
no $I_x$ properly contains another $I_{x'}$ and similarly for the $J_y$'s.)

These results are {\em dichotomies} in the sense that for each $H$ the problem
is polynomial time solvable or NP-complete. They have subsequently been studied
in more general contexts, for graphs with possible loops, for digraphs, and for 
general relational structures (in the context of constraint satisfaction problems). 
In particular, there is a dichotomy for the homomorphism problem for graphs with 
possible loops \cite{hn}, but dichotomy is only conjectured for digraphs (and 
more general structures) \cite{fv,bjk}. A dichotomy for list homomorphism problems 
for graphs with possible loops was established in \cite{fhh,fhh2}, then a general 
dichotomy was proved for all relational systems in \cite{bul}. (A more structural 
dichotomy classification for digraphs was given in \cite{hrsoda}.) For minimum cost
homomorphism problems, a dichotomy for graphs with possible loops is given in
\cite{min-cost}. A structural dichotomy classification for digraphs was conjectured
in \cite{gutin}, and proved in \cite{hr} (cf. \cite{arxiv09,manuscript07}). Then a general
dichotomy for all relational systems was proved in \cite{takh}. Even more general
dichotomy results are known, for so-called finite valued constraint satisfaction 
problems~\cite{zivny}.

It is easy to see that minimum cost homomorphism problems generalize
list homomorphism problems, which in turn generalize homomorphism
problems. Minimum cost homomorphism problems also generalize two
graph optimization problems, the minimum colour sum problem, and the
optimum cost chromatic partition problem \cite{barnoy,supowit}. In the
former, the cost function has only two values, $0$ and $1$ (and $k=0$).
In the latter, the cost function is assumed to be constant across $V(G)$,
i.e., $c(x,u)=c(u)$ for all $x \in V(G)$. This restriction, that costs
only depend on vertices of $H$, appears quite natural even for the general
minimum cost homomorphism problems, and appears not have been studied. In
this paper we take the first steps in investigating its complexity.

Let $H$ be a fixed graph. The {\em minimum constrained cost homomorphism problem} for $H$ has as input 
a graph $G$, together with a cost function $c: V(H) \rightarrow \mathbb{N}$, and an integer $k$, and asks 
whether there is a homomorphism $f : G \to H$ of total cost $cost(f) = \sum_{u \in V(G)}{c(f(u))} \leq k$.

It appears that the added constraint on the cost function may leave the dichotomy classification from
\cite{min-cost} unchanged; in fact, we can show it does not change it for trees $H$ (and in some 
additional cases, cf. Lemma \ref{mchc-biclaw} below.)

\begin{theorem}\label{thm-trees}
Let $H$ be a fixed tree.
Then the minimum constrained cost homomorphism problem to $H$ is polynomial time solvable
if $H$ is a proper interval bigraph, and is NP-complete otherwise.
\end{theorem}

We believe the same may be true for general graphs $H$. We have obtained the following partial classification.

\begin{theorem}\label{thm-cycles}
Let $H$ be a fixed graph.
Then the minimum constrained cost homomorphism problem to $H$ is polynomial time solvable
if $H$ is a proper interval bigraph, and is NP-complete if $H$ is not a chordal bipartite graph.
\end{theorem}

Of course, the first statement of the theorem follows from \cite{min-cost}. Only the second claim,
the NP-completeness, needs to be proved. A bipartite graph $H$ is {\em chordal bipartite} if it
does not contain an induced cycle of length greater than four. Both chordal bipartite graphs and
proper interval bigraphs can be recognized in polynomial time \cite{muller,spinrad}. Proper interval bigraphs 
are a subclass of chordal bipartite graphs, and Lemma 8 below gives a forbidden subgraph 
characterization of proper interval bigraphs within the class of chordal bipartite graphs.  

Our NP-completeness reductions in the proofs of Theorems~\ref{thm-trees} and~\ref{thm-cycles}
use a shorthand, where vertices $v$ of the input graph $G$ have {\em weights} $w(v)$. Adding 
polynomially bounded vertex weights does not affect the time complexity of our problems. Let 
$G, H$ be graphs, and, for every $v \in V(G)$ and every $i \in V(H)$, let $c_i(v)$ denote
the cost of mapping $v$ to $i$. Let $w: V(G) \rightarrow \mathbb{N}$ be a weight function.
The weighted cost of a homomorphism $f: G \to H$ is $cost(f) = \sum_{v \in V(G)} w(v).c_{f(v)}(v)$.
In the {\em weighted minimum cost homomorphism problem} for a fixed graph $H$, the input is
a graph $G$, together with cost functions~$c_i : V(G) \to \mathbb{N}$ (for all $i \in V(H)$), vertex 
weights $w: V(G) \rightarrow \mathbb{N}$, and an integer $k$; and the question is if there is a
homomorphism of $G$ to $H$ of weighted cost at most $k$. 

The variant with constrained costs is defined similarly: the {\em weighted minimum constrained cost 
homomorphism problem} for $H$ has as input a graph $G$, cost function $c: V(H) \rightarrow \mathbb{N}$,
vertex weights $w: V(G) \rightarrow \mathbb{N}$, and an integer $k$, and it
asks if there is a homomorphism $f : G \to H$ with cost $\sum_{v \in V(G)} w(v).c(f(v)) \leq k$.

Clearly, when $w$ is a polynomial function, the weighted minimum cost homomorphism problem and
the minimum cost homomorphism problem are polynomially equivalent. It turns out that this is also the case
for the problems with constrained costs.

\begin{theorem}
\label{mchc:weighted}
Let $H$ be a fixed graph.
The minimum constrained cost homomorphism problem to $H$
and the weighted minimum constrained cost homomorphism problem to $H$ 
with polynomial weights are polynomially equivalent.
\end{theorem}

\section{Chordal Bipartite Graphs}
\label{section}

In this section, we investigate the minimum constrained cost homomorphism problem
for graphs $H$ that contain an even cycle of length at least six as an induced subgraph.
First we treat the case of hexagon, then we handle longer cycles.

\begin{lemma}\label{lemma-hexagon}
Let $H$ be a graph which contains hexagon as an induced subgraph. 
Then, the weighted minimum constrained cost homomorphism problem to $H$ 
is NP-complete.
\end{lemma}

For a fixed graph $H$, the {\em pre-colouring extension problem} to $H$ takes as
input a graph $G$ in which some vertices $v$ have been pre-assigned to images
$f(v) \in V(H)$ (we say $v$ is {\em pre-coloured} by $f(v)$), and asks whether or 
not there exists a homomorphism $f : G \to H$ that extends this pre-assignment. 
This can be viewed a special case of the list
homomorphism problem to $H$ (all lists are either singletons or the entire set $V(H)$),
and has been studied under the name of One-Or-All list homomorphism problem,
denoted OAL-HOM($H$)~\cite{fhh}. Here we adopt the abbreviation 
OAL-HOM($H$) for the pre-colouring extension problem.

The problem OAL-HOM($H$) was first studied in~\cite{feder-98,fhh}.

\begin{lemma}\label{oal:irr-cycle}\cite{fhh}
Let $C$ be a cycle of length $2k$ with $k \geq 3$. 
Then the pre-colouring extension problem to $C$
is NP-complete.
\end{lemma}

We can now present the proof of Lemma~\ref{lemma-hexagon}.

\vskip 2mm
\begin{proof}
The membership in NP is clear. 
Let $C = 1, 2, \cdots, 6$ denote the hexagon and $h_1 h_2 \cdots h_6$ be an induced subgraph of $H$
which is isomorphic to $C$. We reduce from the pre-colouring extension homomorphism problem to $C$.

Let $(G,L)$ be an instance of OAL-HOM($C$), i.e., $G$ is a bipartite graph with $n \geq 2$ vertices and 
$m \geq 1$ edges, and some vertices $v$ of $G$ have been pre-assigned to $f(v) \in V(C)$.
We construct an instance $(G', c, w, T)$ of the weighted minimum constrained cost homomorphism 
problem to $H$ as follows. The graph $G'$ is a bipartite graph obtained from a copy of $G$, by adding, 
for every vertex $v \in V(G)$ pre-coloured $k$, a gadget that is the cartesian product of $v$ and the hexagon,
using six new vertices $(v,1), (v,2), \cdots, (v,6)$, and six new edges $(v,1)(v,2), (v,2)(v,3), \cdots, (v,6)(v,1)$.
We also connect $v$ to exactly two neighbours of $(v,k)$ in its corresponding gadget.
A vertex $v$ and its corresponding gadget is illustrated in Figure~\ref{hexagon-gadget}.

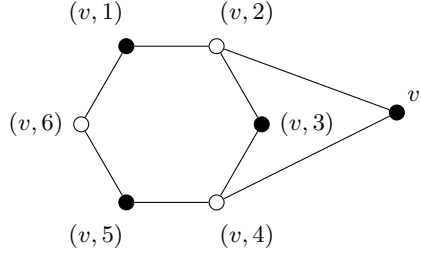
\begin{figure}[ht]
\begin{center}
\begin{tikzpicture}
\tikzstyle{every node} =
[circle, inner sep=2pt, fill=white, draw=black, font=\small]
\foreach \x [] in {2,4,6}
	\node[label=180-60*\x: ${(v,\x)}$] (N\x) at (180-60*\x:1.2) {};
\tikzstyle{every node} =
[circle, inner sep=2pt, fill=black, draw=black, font=\small]
\foreach \x [] in {1,3,5}
	\node[label=180-60*\x:  ${(v,\x)}$] (N\x) at (180-60*\x:1.2) {};
\node[label=45:$v$] (v) at (3: 3) {};

\foreach \j [count = \i] in {2, 3, 4, 5, 6, 1}
	\draw (N\i) -- (N\j) ;
\draw (v) -- (N2);
\draw (v) -- (N4);
\end{tikzpicture}
\end{center}
\caption{A gadget in $G'$ for a vertex $v \in V(G)$ pre-coloured by $3$}
\label{hexagon-gadget}
\end{figure}

We define the vertex weight function $w$ as follows.
\begin{itemize}
\item for every vertex $v$ in the copy of $G$, let $w(v) = 1$
\item for every pre-coloured vertex $v \in V(G)$:
	\begin{itemize}
	\item $w((v,1)) = w((v,4)) = 5 \times 36n^3 + 1$,
	\item $w((v,2)) = w((v,5)) = 1$,
	\item $w((v,3)) = 36n^2$,
	\item $w((v,6)) = 6n$.
	\end{itemize} 
\end{itemize}

We define the homomorphism cost function $c$ as follows.
\begin{itemize}
\item $c(h_1) =  c(h_4) = 0$,
\item $c(h_2) =  c(h_5) = 36n^2$,
\item $c(h_3) = 1$,
\item $c(h_6) = 6n$,
\item $c(h_i) = 5\times 36n^3 + 1$ for all other vertices $h_i \in V(H)$.
\end{itemize}

Finally, we set $T = 5 \times 36n^3 = 180n^3$.

We now claim that there is an extension of the pre-colouring $f$ to a 
homomorphism of $G$ to $C$ if and only if there is a homomorphism 
of $G'$ to $H$ with weighted cost at most $T$. 

First, assume that the pre-colouring can be extended to a homomorphism
$f : G \to C$. We define a homomorphism $g : G' \to H$ as follows.

\begin{itemize}
\item $g(u) = h_i$ iff $f(u) = i$ for every vertex $u \in V(G)$ and every $1 \leq i \leq 6$,
\item $g((u,i)) = h_i$ for every vertex $u \in V(G)$ pre-coloured $k$ and every $1 \leq i \leq 6$.
\end{itemize}

{\bf Claim}. The function $g$ is a homomorphism of $G'$ to $H$.
Moreover, it only maps vertices of $G'$ to the copy of $C$ in $G$,
i.e., $g$ only uses vertices $h_1, h_2, \cdots, h_6$.
\vskip 2mm

To prove the above claim, we distinguish three types of edges in $G'$.
\begin{enumerate}
\item Edges $uv$ corresponding to the edges in $G$ ($u,v \in V(G)$):
These are clearly mapped to edges in $H$ by $g$ as $g(u) = f(u)$ for
all vertices $u \in V(G)$ and $f$ is a homomorphism of $G$ to $C$.

\item Edges $(u,i)(u,i+1)$ that connect two vertices of the gadgets:
These edges map to the corresponding edge $h_i h_{i+1}$ by
definition of $g$ (indices modulo $6$).

\item Edges that connect a vertex $u \in V(G)$ to two vertices
in its corresponding gadget: 
Notice that there is a gadget for $u$ in $G'$ only when $u$ is 
pre-coloured $i$. So, we have $f(u) = i$.
This further implies that $g(u) = h_i$. Also, notice that $g((u,i-1)) = h_{i-1}$
and $g((u,i+1)) = h_{i+1}$ by the definition of $g$ (again, all indices modulo $6$).
Hence, edges $u(u,i-1)$ and $u(u,i+1)$ also map to edges $h_{i-1} h_i$ and $h_i h_{i+1}$, respectively.
\end{enumerate}

This completes the proof of the above Claim. We now show that the cost of $g$
is at most $T = 180n^3$.

\begin{itemize}
\item For every vertex $u \in V(G)$, $w(u) = 1$ and $c(g(u)) \leq 36n^2$.
Also, there are exactly $n$ such vertices in $G'$. This contributes at most $36n^3$
to the cost of the homomorphism.

\item For every pre-coloured vertex $u \in V(G)$, its corresponding gadget
contributes exactly $4\times 36n^2$:
	\begin{itemize}
	\item vertices $(u,1)$ and $(u,4)$ do not contribute, as $c(h_1) = c(h_4) = 0$,
	\item vertices $(u,2)$ and $(u,5)$ each contribute $36n^2$,
	\item vertices $(u,3)$ and $(u,6)$ each contributes $36n^2 = 6n\times 6n = 36n^2 \times 1$.
	\end{itemize}
\end{itemize}
There are at most $n$ gadgets in $G'$ (one for every vertex $u \in V(G)$), and so,
the total contribution of all vertices of the gadgets is at most $4 \times 36n^3$.
Therefore, the cost of $g$ is at most $5 \times 36n^3 = 180n^3 = T$.

Conversely, let $g$ be a homomorphism of $G'$ to $H$ which costs at most $T$.
We prove that there is a homomorphism $f : G \to C$ extending the pre-colouring.
First, we show that $g$ has the following two properties.

\begin{itemize}
\item It only maps vertices of $G'$ to the vertices of the hexagon
$h_1, h_2 ,\cdots, h_6$, 

\item all gadgets are mapped identically to the hexagon in $H$,
that is, for all pre-coloured vertices $u \in V(G)$ and for every 
$1 \leq i \leq 6$, $g((u,i)) = h_i$.
\end{itemize}

The first property holds because $c(a) > T$ for every vertex $a \in V(H)$
other than the vertices of the hexagon (and the fact that, by definition,
all vertex weights are positive integers).
In fact, we must have $w(u) \times c(g(u)) \leq T$, or equivalently,
$c(g(u)) < {(T+1) \over w(u)}$, for every vertex $u \in V(G')$.
This restricts possible images of vertices with large vertex weights.
Consider vertices in the gadget of a vertex $u \in V(G')$. For instance,
every $(u,4)$ must map to either $h_1$ or $h_4$. Similarly, none of the
$(u,3)$ vertices can map to any vertex other than $h_1$, $h_3$, or $h_4$.
Given that $(u,3)$ and $(u,4)$ are adjacent in $G'$, their images must
also be adjacent in $H$. This enforces $f((u,3)) = h_3$ and $f((u,4)) = h_4$
(for every $u$ that has a gadget in $G'$). Similar to $(u,4)$, 
$g$ must also map every $(u,1)$ to either $h_1$ or $h_4$, but $g((u,1)) = h_4$
is not feasible as it does not leave any options for the image of $(u,2)$.
Hence, $g((u,1)) = h_1$. This further implies that $g((u,6)) = h_6$ (as it is adjacent
to $(u,1)$), and finally, $g((u,2)) = h_2$ and $g((u,5)) = h_5$.

It is now easy to verify that for every vertex $u \in V(G)$ pre-coloured $j$,
we always have $g(u) = h_j$. This is because $u$ is adjacent to 
$(u,j-1)$ and $(u,j+1)$ in $G'$ and the only vertex in $H$ that is adjacent to
the $g((u,j-1)) = h_{j-1}$ and $g((u,j+1)) = h_{j+1}$ and the cost of mapping 
to it is less than or equal to $T$ is $h_j$. This completes
the proof as we can define a homomorphism $f : G \to C$ extending the
pre-colouring by setting $f(v) = i \iff g(v) = h_i$.
\end{proof}

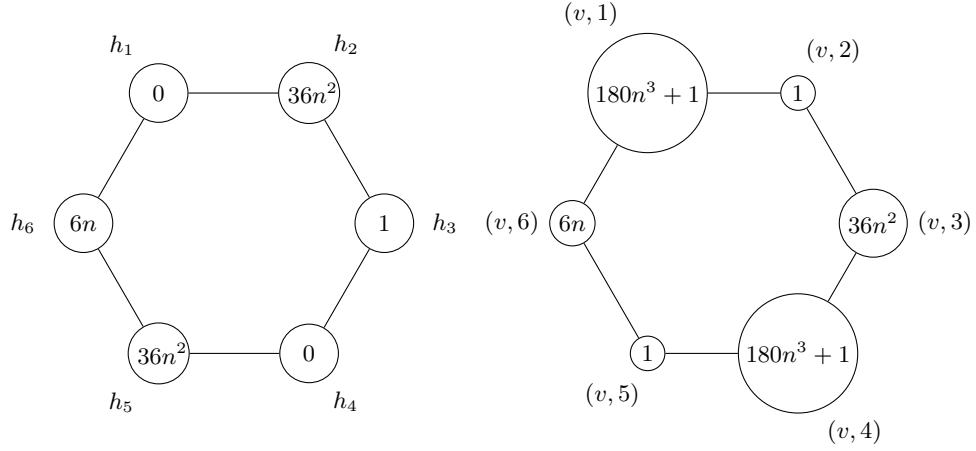
\begin{figure}[h]
\begin{center}
\begin{tikzpicture}

\begin{scope}[shift={(-1,0)},local bounding box=aa]
\tikzstyle{every node} =
[circle, inner sep=2pt, fill=white, draw=black, text centered, text width=16, font=\small]
\foreach \x [count = \i] in {$0$, $36n^2$, $1$, $0$, $36n^2$, $6n$}
	\node[label=180-60*\i: $h_\i$] (N\i) at (180-60*\i:2) {\x};
\foreach \j [count = \i] in {2, 3, 4, 5, 6, 1}
	\draw (N\i) -- (N\j) ;
\end{scope}

\begin{scope}[shift={(5.5,0)},local bounding box=bb]
\tikzstyle{every node} =
[circle, inner sep=2pt, fill=white, draw=black, text centered, font=\small]
\foreach \x [count = \i] in {$180n^3 + 1$, $1$, $36n^2$, $180n^3 + 1$, $1$, $6n$}
	\node[label=180-60*\i: ${(v,\i)}$] (N\i) at (180-60*\i:2) {\x};
\foreach \j [count = \i] in {2, 3, 4, 5, 6, 1}
	\draw (N\i) -- (N\j);
\end{scope}

\end{tikzpicture}
\end{center}
\caption{A hexagon in $H$ together with associated homomorphism costs (left),
and a gadget in $G'$ together with vertex weights (right).}
\label{hexagon-short}
\end{figure}

\vskip 2mm
A shorthand of the construction used in the above proof 
is shown in Figure~\ref{hexagon-short}.

We now extend Lemma~\ref{lemma-hexagon} to larger even cycles.

\begin{lemma}\label{lemma-cycles}
Let $H$ be a bipartite graph which contains a cycle of length at least eight
as an induced subgraph. 
Then the weighted minimum constrained cost homomorphism problem to $H$
is NP-complete.
\end{lemma}

{\bf Proof sketch}.
The proof is similar to the proof of Lemma~\ref{lemma-hexagon}. We only discuss the reduction here.
Let $C = 1, 2, \cdots, {2k}$ be an even cycle, and $h_1 h_2 \cdots h_{2k}$ be an induced subgraph of $H$
which is isomorphic to $C$ ($k \geq 4$). Again, we reduce from OAL-HOM($C$).
We take an instance of the OAL-HOM($C$), i.e., a graph $G$ with $n \geq 2$ vertices and $m \geq 1$ edges,
with some vertices of $G$ pre-coloured by vertices of $C$. We construct a corresponding
instance $(G', c, w, T)$ of the weighted minimum constrained cost homomorphism problem to $H$.

The graph $G'$ is constructed exactly as before: we start with a copy of $G$ and for every vertex $v$
pre-coloured by $t$, we add the cartesian product of $v$ and $C$ using $2k$ new vertices and $2k$ 
new edges. Finally, make $v$ adjacent to two vertices in its corresponding gadget, $(v,t-1)$ and $(v,t+1)$ 
(all indices modulo $2k$).

We define the vertex weight function $w$ as follows.
\begin{itemize}
\item for every vertex $v$ in the copy of $G$, let $w(v) = 1$ 
\item for every pre-coloured vertex $v \in V(G)$:
	\begin{itemize}
	\item $w((v,1)) = w((v,4)) = 50kn^2$,
	\item $w((v,2)) = w((v,3)) = w((v,5)) = 1$,
	\item $w((v,i)) = 9n$ for all $6 \leq i \leq 2k$
	\end{itemize} 
\end{itemize}

We define the homomorphism cost function $c$ as follows. 
\begin{itemize}
\item $c(h_1) =  c(h_4) = 0$,
\item $c(h_2) =  c(h_3) = c(h_5) = 8kn$,
\item $c(h_i) = 1$ for all $6 \leq i \leq 2k$,
\item $c(h_i) = 50kn^2$ otherwise.
\end{itemize}

Finally, we set $T = 50kn^2 - 1$. As in the proof of Lemma~\ref{lemma-hexagon},
we argue that there is a homomorphism of $G$ to $C$ extending the pre-colouring
if and only if there is a homomorphism of $G'$ to $H$ with cost at most $T$.

This completes the proof of Theorem \ref{thm-cycles}, as chordal bipartite graphs
have no induced cycles of length greater than four.



%

We note that Theorem~\ref{thm-cycles} gives only a partial dichotomy for
the minimum constrained cost homomorphism problem,
as there is a gap between the class of chordal bipartite graphs and
the class of proper interval bigraphs. Specifically, the following result
clarifies the gap.

\begin{lemma}\cite{interval-circular}\label{here}
\label{pib-forbid}
A chordal bipartite graph $H$ is a proper interval bigraph if and only if it does not contain
a bipartite claw, a bipartite net, or a bipartite tent.
\end{lemma}


\begin{center}
\begin{figure}[h]
\begin{center}
\begin{tabular}{c@{\hspace{3em}}c@{\hspace{3em}}c}
\begin{tikzpicture}
\node at (0.0, 0.0) (a) {};
\node at (0.0, 2.0) (b) {};
\node at (2.0, 2.0) (c) {};
\node at (2.0, 0.0) (d) {};
\node at (0.0, 1.0) (x) {};
\node at (1.0, 1.0) (y) {};
\node at (1.0, 0.0) (z) {};
\draw[thick] (a) -- (x) -- (b);
\draw[thick] (a) -- (y) -- (c);
\draw[thick] (a) -- (z) -- (d);
\foreach \t in {(a),(b),(c),(d)}
	\draw[fill=black] \t circle (0.15);
\foreach \t in {(x),(y),(z)}
	\draw[fill=white] \t circle (0.15);
\end{tikzpicture}
&
\begin{tikzpicture}
\node at (0.0, 1.0) (a) {};
\node at (1.0, 0.0) (b) {};
\node at (2.0, 1.0) (c) {};
\node at (1.0, 2.0) (d) {};
\node at (1.0, 1.0) (x) {};
\node at (0.0, 0.0) (y) {};
\node at (2.0, 0.0) (z) {};
\draw[thick] (a) -- (x) -- (c) -- (z) -- (b) -- (y)  -- (a);
\draw[thick] (b) -- (x) -- (d);
\foreach \t in {(x),(y),(z)}
	\draw[thick] (b) -- \t;
\foreach \t in {(a),(b),(c),(d)}
	\draw[fill=black] \t circle (0.15);
\foreach \t in {(x),(y),(z)}
	\draw[fill=white] \t circle (0.15);
\end{tikzpicture}
& 
\begin{tikzpicture}
\node at (0.0, 0.0) (a) {};
\node at (1.0, 1.0) (b) {};
\node at (2.0, 0.0) (c) {};
\node at (0.0, 2.0) (d) {};
\node at (0.0, 1.0) (x) {};
\node at (1.0, 0.0) (y) {};
\node at (2.0, 1.7) (z) {};
\draw[thick] (a) -- (y) -- (c);
\draw[thick] (a) -- (x) -- (d);
\foreach \t in {(x),(y),(z)}
	\draw[thick] (b) -- \t;
\foreach \t in {(a),(b),(c),(d)}
	\draw[fill=black] \t circle (0.15);
\foreach \t in {(x),(y),(z)}
	\draw[fill=white] \t circle (0.15);
\end{tikzpicture}
\end{tabular}
\end{center}
\caption{The bipartite claw, net and tent}
\label{fig-min-max}
\end{figure}
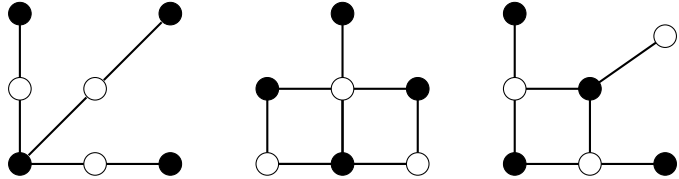
\end{center}

\section{The Dichotomy For Trees}
\label{5:sec:claw}

In this section, we prove an extension of Theorem~\ref{thm-cycles} to graphs $H$ that
contain a bipartite claw. As in the case of large cycles, we focus on the
weighted version of the problem and show that it is NP-complete when the
target graph $H$ contains a bipartite claw. As a corollary we will obtain our dichotomy 
classification for trees, Theorem \ref{thm-trees}.

\begin{lemma}\label{mchc-biclaw}
Let $H$ be a fixed graph containing the bipartite claw as an induced subgraph.
Then the weighted minimum constrained cost homomorphism problem to $H$
is NP-complete.
\end{lemma}

It is well known that the problem of finding a maximum independent set in a 
graph is NP-complete. Alekseev and Lozin~cite{lozin} proved
that the problem is still NP-complete even when the input is restricted
to be a $3$-partite graph, cf. Gutin, Hell, Rafiey and Yeo~\cite{min-cost}.

\begin{theorem}\label{mis-3p}\cite{lozin,min-cost}
The problem of finding a maximum independent set in a $3$-partite graph $G$,
even given the three partite sets, in NP-complete.
\end{theorem}

The main idea of the proof of Lemma~\ref{mchc-biclaw} is similar to the
proofs of Lemmata~\ref{lemma-hexagon} and~\ref{lemma-cycles}.
We show that finding an independent set of size at least $k$ in an arbitrary 
$3$-partite graph $G$ is equivalent to finding a homomorphism of cost 
at most $k'$ in an auxiliary graph $G'$ together with constrained costs $c$ and
vertex weights~$w$. To construct $G'$, we start by adding a fixed number of
placeholder vertices; vertices that, with the appropriate weights and costs,
always map to the same specific vertices of the target graph $H$ in any homomorphism of 
$G'$ to $H$ of minimum cost.
We then use these placeholder vertices in our construction to ensure that the vertices corresponding to
each part of the the input graph $G$ are only mapped to certain vertices of $H$.


\vskip 2mm
\begin{proof}
The membership in NP is clear. To show that the problem is NP-hard, we reduce
from the problem of finding a maximum independent set in a $3$-partite graph,
stated in Theorem~\ref{mis-3p}.
Let $G$ be a a $3$-partite graph in which we seek an independent set of size $k$,
with parts $V_1$, $V_2$, and $V_3$, and denote by and $n$ and $m$ the number
of vertices and edges in $G$, respectively.
We assume that $G$ is non-empty. Without loss of generality, we can assume that $|V_1| \geq 1$.
We construct an instance $(G', c, w, T_{G,k})$ of the weighted minimum cost
graph homomorphism and show that $G$ has an independent set of size $k$
if and only if there is a homomorphism of $G'$ to $H$ with cost less
than or equal to $T_{G,k}$.

We construct the bipartite graph $G'$ as follows. 
Subdivide every edge $e$ in $G$ using a new vertex $d_e$
(which is adjacent to both ends of $e$). Add three vertices
$b_1$, $b_2$ and $b_3$ and make each $b_i$ adjacent to all
vertices in $V_i$  for $i=1,2,3$. Finally, add three more
vertices $c_0$, $c_1$ and $c_2$. Make $c_0$ adjacent to
$b_1$, $b_2$ and $b_3$, $c_1$ adjacent to $b_1$ and $c_2$
adjacent to $b_2$. A $3$-partite graph $G$ together with
its corresponding $G'$ is depicted in Figure~\ref{fig-biclaw-gadget}.
For future reference, we denote the set 
$\{ b_1, b_2, b_3, c_0, c_1, c_2 \}$ by $V_4$.

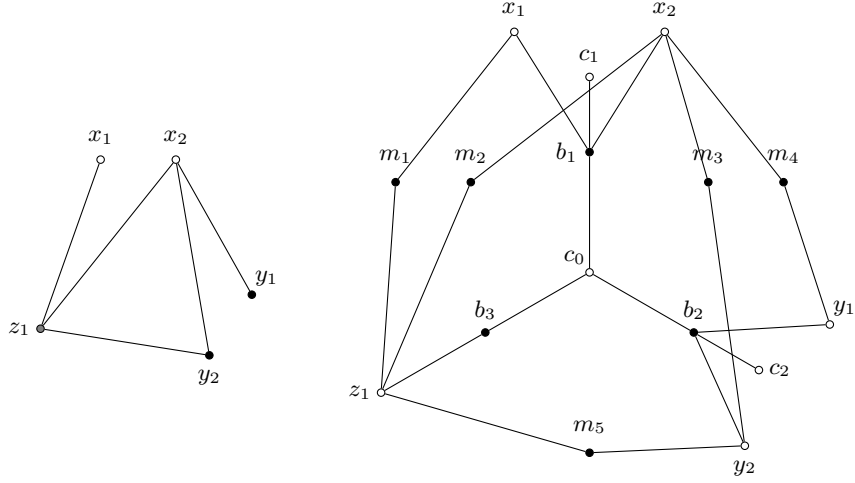
\begin{figure}[ht]
\begin{center}
\begin{tikzpicture}

\begin{scope}[local bounding box=aa]
\tikzstyle{every node} =
[circle, inner sep=1pt, fill=white, draw=black, font=\small]
\node[label=90:$x_1$] (X1) at ([shift=(0:-0.5)]90:1.5) {};
\node[label=90:$x_2$] (X2) at ([shift=(0:0.5)]90:1.5) {};

\tikzstyle{every node} =
[circle, inner sep=1pt, fill=black, draw=black, font=\small]
\node[label=60:$y_1$] (Y1) at ([shift=(65:0.5)]-30:1.5) {};
\node[label=-90:$y_2$] (Y2) at ([shift=(45:-0.5)]-30:1.5) {};

\tikzstyle{every node} =
[circle, inner sep=1pt, fill=gray, draw=black, font=\small]
\node[label=180:$z_1$] (Z1) at (210:1.5) {};

\draw (X1) -- (Z1) -- (X2);
\draw (X2) -- (Y1);
\draw (X2) -- (Y2) -- (Z1);
\end{scope}


\begin{scope}[shift={(6,0)},local bounding box=bb]

\tikzstyle{every node} =
[circle, inner sep=1pt, fill=white, draw=black, font=\small]

\node[label=90:$x_1$] (X1) at ([shift=(0:-1)]90:3.2) {};
\node[label=90:$x_2$] (X2) at ([shift=(0:1)]90:3.2) {};

\node[label=60:$y_1$] (Y1) at ([shift=(65:1)]-30:3.2) {};
\node[label=-90:$y_2$] (Y2) at ([shift=(45:-1)]-30:3.2) {};

\node[label=180:$z_1$] (Z1) at (210:3.2) {};

\node[label=135:$c_0$] (C0) at (0,0) {};
\node[label=90:$c_1$] (C1) at (90:2.6) {};
\node[label=0:$c_2$] (C2) at (-30:2.6) {};

\tikzstyle{every node} =
[circle, inner sep=1pt, fill=black, draw=black, font=\small]

\node[label=180:$b_1$] (B1) at (90:1.6) {};
\node[label=90:$b_2$] (B2) at (-30:1.6) {};
\node[label=90:$b_3$] (B3) at (210:1.6) {};

\node[label=90:$m_1$] (M1) at ([shift=(0:-0.5)]150:2.4) {};
\node[label=90:$m_2$] (M2) at ([shift=(0:0.5)]150:2.4) {};

\node[label=90:$m_3$] (M3) at ([shift=(0:-0.5)]30:2.4) {};
\node[label=90:$m_4$] (M4) at ([shift=(0:0.5)]30:2.4) {};

\node[label=90:$m_5$] (M5) at (-90:2.4) {};

\draw (C1) -- (B1) -- (C0) -- (B3);
\draw (C2) -- (B2) -- (C0);

\draw (X2) -- (B1) -- (X1);
\draw (Y2) -- (B2) -- (Y1);
\draw (Z1) -- (B3);

\draw (X1) -- (M1) -- (Z1) -- (M2) -- (X2);
\draw (X2) -- (M4) -- (Y1);
\draw (X2) -- (M3) -- (Y2) -- (M5) -- (Z1);
\end{scope}

\end{tikzpicture}
\end{center}
\caption{A $3$-partite graph $G$ with parts $V_1 = \{x_1, x_2\}$, $V_2 = \{ y_1, y_2\}$, $V_3 = \{ z_1\}$ (left) 
and its corresponding bipartite graph $G'$ (right)}
\label{fig-biclaw-gadget}
\end{figure}

Let $H'=(X,Y)$ be an induced subgraph of $H$ which is isomorphic to
a bipartite claw with parts $X = \{v_0, v_1, v_2, v_3 \}$
and $Y = \{ u_1, u_2, u_3\}$, and edge set
$$E' = \{ u_1 v_1, u_2 v_2, u_3 v_3, u_1 v_0, u_2 v_0, u_3 v_0\}.$$

Define the homomorphism cost function $c$ as follows (see Figure~\ref{fig-biclaw}).
\begin{itemize}
\item $c(v_0) = 4$
\item $c(v_1) = c(u_1) = 1$
\item $c(u_2) = c(v_3) = 3$
\item $c(v_2) = c(u_3) = 0$
\item $c(u) = 160n(m+n)$ for every other vertex $u \notin X \cup Y$ 
\end{itemize}

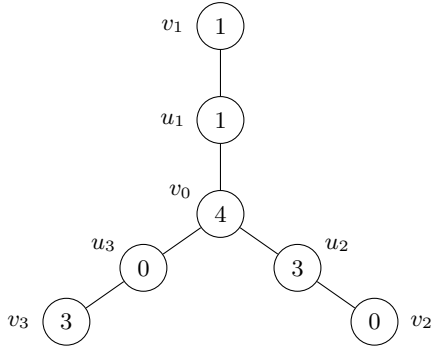
\begin{figure}[ht]
\begin{center}
\begin{tikzpicture}
\tikzstyle{every node} =
[circle, inner sep=1pt, fill=white, draw=black, text centered, text width=14, font=\small]

\node[label=165:$v_0$] (V0) at (0, 0) {$4$};

\node[label=180:$v_1$] (V1) at (90:2.5) {$1$};
\node[label=0:$v_2$] (V2) at (-35:2.5) {$0$};
\node[label=180:$v_3$] (V3) at (-145:2.5) {$3$};

\node[label=180:$u_1$] (U1) at (90:1.25) {$1$};
\node[label=15:$u_2$] (U2) at (-35:1.25) {$3$};
\node[label=165:$u_3$] (U3) at (-145:1.25) {$0$};

\draw (V0) -- (U1) -- (V1);
\draw (V0) -- (U2) -- (V2);
\draw (V0) -- (U3) -- (V3);
\end{tikzpicture}
\end{center}
\caption{A bipartite claw, with homomorphism costs}
\label{fig-biclaw}
\end{figure}

Define the vertex weights of $G'$ as follows.
\begin{itemize}
\item $w(b_1) = w(c_1) = 50n(m+n)$
\item $w(b_3) = w(c_2) = 160n(m+n)$
\item $w(b_2) = w(c_0) = 1$
\item $w(u) = 4(m+n)$ for every vertex $u \in V_1$
\item $w(u) = 3(m+n)$ for every vertex $u \in V_2$
\item $w(u) = 12(m+n)$ for every vertex $u \in V_3$
\end{itemize}

Finally, let $T_{G,k}$ be the sum of the following values.
\begin{itemize}
\item $T^1_{G,k} = 16(m+n)|V_1|$,
\item $T^2_{G,k} = 12(m+n)|V_2|$,
\item $T^3_{G,k} = 48(m+n)|V_3|$,
\item $T^4_{G,k} = 2 \times 50n(m+n) + 4 + 3$, 
\item $T^e_{G,k} = 3m$, and,
\item $T^I_{G,k} = -12(m+n)k$.
\end{itemize}

Or equivalently:
$$T_{G,k} = 100n(m+n) + 7 + 3m + (4|V_1|+36|V_3|)(m+n)+ 12(m+n)(n-k)$$
We prove that $G$ has an independent set of size $k$ if and only if
there is a homomorphism of $G'$ to $H$ of cost less than or equal to $T_{G,k}$.

First, assume that $I$ is an independent set of size $k$ in $G$
with parts $I_1 \subset V_1$, $I_2 \subset V_2$, and $I_3 \subset V_3$.
Let $k_i$ denote $|I_i|$ ($i=1,2,3$).
Define the homomorphism $f_I$ as follows.
\begin{itemize}
\item $f_I(u) = v_i$ for all vertices $u \in I_i$ ($i = 1, 2, 3$),
\item $f_I(u) = v_0$ for all vertices $u \in V(G) - I$,
\item $f_I(d_e) = u_j$ for every edge $e$ with one end in $I_j$ ($j = 1,2,3$),
\item $f_I(d_e) = u_3$ for every edge $e$ with both ends in $V - I$,
\item $f_I(b_j) = u_j$ for $j = 1,2,3$, and finally,
\item $f_I(c_k) = v_k$ for $k = 0, 1, 2$.
\end{itemize}

Notice that at most one end of each edge is in $I$, hence,
the above assignment is indeed a function. In fact, it is easy to verify that $f_I$
is a homomorphism. 
\begin{itemize}
\item edges subdivided from edges $e$ with both ends in $V-I$ map to $v_0 u_3$,
\item edges subdivided from edges $e$ with one end in $I_i$ and the other end in $V-I$
map to $u_i v_i$ and $u_i v_0$ ($i=1,2,3$),
\item edges connecting $b_i$ to $V_i$ map to $u_i v_i$ ($i=1,2,3$),
\item $c_0 b_i$ map to $v_0 u_i$ ($i=1,2,3$), and,
\item $b_i c_i$ map to $v_i u_i$ ($i=1,2$).
\end{itemize}

We now compute the cost of $f_I$ and show that it does not exceed $T_{G,k}$.
\begin{itemize}
\item The vertices in $V_1$ contribute exactly $(|V_1| - k_1) \times 16(m+n) + k_1 \times 4(m+n)$, or,
$T^1_{G,k} - 12k_1(m+n)$,

\item the vertices in $V_2$ contribute exactly $(|V_2| - k_2) \times 12(m+n) + k_1 \times 0$, or,
$T^2_{G,k} - 12k_2(m+n)$,

\item the vertices in $V_3$ contribute exactly $(|V_3| - k_3) \times 48(m+n) + k_3 \times 36(m+n)$, or,
$T^3_{G,k} - 12k_3(m+n)$,

\item the vertices in $V_4$ contribute a total of $100n(m+n) + 7 = T^4_{G,k}$ (see Table~\ref{claw-contribution-v4}),

\item the vertices $d_e$ contribute at most $3m = T^e_{G,k}$.
\end{itemize}

Notice that $k = k_1 + k_2 + k_3$, hence, the cost of $f_I$ is at most $T_{G,k}$.

\begin{table}[ht]
\begin{center}
\begin{tabular}{|c|c|c|c|c|}
\hline
vertex $v$   &  $w(v)$       & $f_I(v)$ & $c(f_I(v))$ & contributed cost of $v$ \\ \hline
$b_1$        &  $50n(m+n)$   & $u_1$    & $1$ & $50n(m+n)$  \\
$b_2$        &  $1$          & $u_2$    & $3$ & $3$  \\
$b_3$        &  $160n(m+n)$  & $u_3$    & $0$ & $0$  \\
$c_0$        &  $1$          & $v_0$    & $4$ & $4$  \\
$c_1$        &  $50n(m+n)$   & $v_1$    & $1$ & $50n(m+n)$  \\
$c_2$        &  $160n(m+n)$  & $v_2$    & $0$ & $0$  \\
\hline
\end{tabular}
\end{center}
\caption{contribution of vertices in $V_4$ to the cost of homomorphism $f_I$}
\label{claw-contribution-v4}
\end{table}

Conversely, assume that $f$ is a homomorphism of $G'$ to $H$ which
costs less than or equal to $T_{G,k}$. Note that $T_{G,k} < 150n(m+n)$. 
This prevents any vertex $v$ to map to a vertex $a$ when
$c(v,a) \times w(v) \geq T_{G,k}$. In particular, $b_1$ and $c_1$ can only map 
to vertices $a$ with $c(a) < 3$, i.e, $v_1, u_1, v_2, u_3$.
But $b_1$ and $c_1$ are adjacent and the only edge in $H$
among these four vertices is $u_1 v_1$.
Similarly, $b_3$ and $c_2$ can only map to $u_3$ or $v_2$. 
Observe that $f(b_3) = v_2$ is not feasible, as it implies
$f(c_0) = u_2$ and hence $f(b_1) \in \{ v_0, v_2 \}$.
Thus, we have $f(b_3) = u_3$, $f(b_1) = u_1$, $f(c_1) = v_1$,
$f(c_0) = v_0$, $f(c_2) = v_2$, and finally $f(b_2) = u_2$.

This restricts possible images of vertices in $V$. 
Specifically, all vertices in $V_1$ are adjacent to $b_1$,
thus, $f$ can only map them to $v_1$ or $v_0$, the neighbourhood of
$u_1 = f(b_1)$. Similarly, each vertex in $V_2$ will only map to
$v_2$ or $v_0$, and each vertex in $V_3$ will only map to $v_3$ or $v_0$.

Let $I$ denote the set of vertices of $G$ that $f$ maps to $v_1$, $v_2$ or $v_3$.
Notice that $I$ is an independent set in $G$. This is because
any two adjacent vertices in $G$ are of distance two in $G'$ but
the shortest path between $v_1$ and $v_2$, or between $v_2$ and $v_3$,
or between $v_3$ and $v_1$ in $H'$ has length $4$.

We complete the proof by showing that $|I| \geq k$. 
Let $|I| = k'$ and assume for a contradiction that $k' < k$.
Let $f_I$ denote the homomorphism of $G'$ to $H$ constructed from
$I$ as described in the first part of the proof with $cost(f_I) \leq T_{G,k'}$.
Observe that $f$ and $f_I$ are identical for every vertex $v \in V_i$ ($i=1,2,3,4$).
Hence, $|cost(f) - cost(f_I)| \leq 3m$. 
This implies that $cost(f_I) \leq cost(f) + 3m$. Also, note that
$cost(f_I) \geq T_{G,k'} - 3m$, hence, we have $T_{G,k'} - 3m \leq T_{G,k} + 3m$,
or equivalently, $T_{G,k'} - T_{G,k} \leq 6m$.
But this is a contradiction because:
$$T_{G,k'} - T_{G,k} = T^I_{G,k'} - T^I_{G,k} = 12(m+n)(k-k') \geq 12(m+n).$$
\end{proof}


We can now apply Theorem~\ref{mchc:weighted} and derive the same conclusion
for the problem without vertex weights.

\begin{theorem}\label{mchc:biclaw-thm}
Let $H$ be a fixed graph containing the bipartite claw as an induced subgraph.
Then the minimum constrained cost homomorphism problem to $H$ is NP-complete.
\end{theorem}

Note that Lemma~\ref{pib-forbid} implies that for trees, a chordal bipartite $H$ is a proper interval 
bigraph if and only if it does not contain an induced bipartite claw. Thus we obtain Theorem
\ref{thm-trees} as a corollary.

%
%
%
%
%
%

\section{Conclusion}

We left open the complexity of the minimum constrained cost graph homomorphism problems
in general. In particular, it remains to check whether the problem is NP-complete also for graphs 
$H$ that contain a bipartite net or a bipartite tent.



\end{document}